\documentclass[sn-basic]{sn-jnl}
\usepackage{graphicx}%
\usepackage{multirow}%
\usepackage{amsmath,amssymb,amsfonts}%
\usepackage{amsthm}%
\usepackage{mathrsfs}%
\usepackage[title]{appendix}%
\usepackage{xcolor}%
\usepackage{textcomp}%
\usepackage{manyfoot}%
\usepackage{booktabs}%
\usepackage{algorithm}%
\usepackage{algorithmicx}%
\usepackage{algpseudocode}%
\usepackage{listings}%
\usepackage{bm}

\newcommand{\fig}[1]{Fig.~\ref{#1}}
\newcommand{\eq}[1]{Eq.~(\ref{#1})}

\begin{document}

\title[Article Title]{Parallel Ising Annealer via Gradient-based Hamiltonian Monte Carlo}


\author[1]{\fnm{Hao} \sur{Wang}}\email{auhitayan@mail.scut.edu.cn}
\equalcont{These authors contributed equally to this work.}

\author[2]{\fnm{Zixuan} \sur{Liu}}\email{21s053071@stu.hit.edu.cn}
\equalcont{These authors contributed equally to this work.}

\author[3]{\fnm{Zhixin} \sur{Xie}}\email{22132120@zju.edu.cn}

\author[3]{\fnm{Langyu} \sur{Li}}\email{langyuli@zju.edu.cn}

\author*[2]{\fnm{Zibo} \sur{Miao}}\email{miaozibo@hit.edu.cn}

\author*[1]{\fnm{Wei} \sur{Cui}}\email{aucuiwei@scut.edu.cn}

\author*[3]{\fnm{Yu} \sur{Pan}}\email{ypan@zju.edu.cn}

\affil[1]{\orgdiv{School of Automation Science and Engineering}, \orgname{ South China University of
Technology}, \orgaddress{\street{381 Wushan Road}, \city{Guangzhou}, \postcode{510641}, \state{Guangdong}, \country{China}}}

\affil[2]{\orgdiv{School of Mechanical Engineering and Automation}, \orgname{Harbin Institute of
Technology}, \orgaddress{\street{Taoyuan Street}, \city{Shenzhen}, \postcode{518055}, \state{Guangdong}, \country{China}}}

\affil[3]{\orgdiv{College of Control Science and Engineering}, \orgname{Zhejiang University}, \orgaddress{\street{38 Zheda
Road}, \city{Hangzhou}, \postcode{310027}, \state{Zhejiang}, \country{China}}}


\abstract{Ising annealer is a promising quantum-inspired computing architecture for combinatorial optimization problems. In this paper, we introduce an Ising annealer based on the Hamiltonian Monte Carlo, which updates the variables of all dimensions in parallel. The main innovation is the fusion of an approximate gradient-based approach into the Ising annealer which introduces significant acceleration and allows a portable and scalable implementation on the commercial FPGA. Comprehensive simulation and hardware experiments show that the proposed Ising annealer has promising performance and scalability on all types of benchmark problems when compared to other Ising annealers including the state-of-the-art hardware. In particular, we have built a prototype annealer which solves Ising problems of both integer and fraction coefficients with up to 200 spins on a single low-cost FPGA board, whose performance is demonstrated to be better than the state-of-the-art quantum hardware D-Wave 2000Q and similar to the expensive coherent Ising machine. The sub-linear scalability of the annealer signifies its potential in solving challenging combinatorial optimization problems and evaluating the advantage of quantum hardware.}




\maketitle

\section{Introduction}\label{sec1}

As a novel computing paradigm, quantum computing has the potential to vastly surpass its classical counterparts in solving challenging problems \cite{Shor1994prime,Grover1996database}. In particular, there may exist efficient quantum algorithms to solve the NP-hard combinatorial optimization problems \cite{Farhi2014QAOA,Albash2018AQC,Guerreschi2019QAOA,Yan2022naqc}. For example, quantum approximation optimization algorithm (QAOA) and quantum annealing (QA) may help us reach the approximate solution faster \cite{Albash2018advan,Mandra2018speedup}.

Quantum annealers require precise hardware and environment control, which are expensive and may not compatible with near-term quantum devices. Inspired by general quantum computing, many models of optical Ising machines and FPGA-based annealers have been proposed for combinatorial optimization \cite{Wang2013CIM,Babaeian2019CIM,Mohseni2022solver, Waidyasooriya2021psqa, Aadit2022massive, lu2023speed}, including the famous optical coherent Ising machine (CIM) which treats the discrete Ising spins as continuous optical phases \cite{Chou2019wim, Vaidya2022oim, honjo2021100}. Ising annealers implemented with the classical computing can also achieve a speedup based on hardware parallelism \cite{Aadit2022massive, Okuyama2019ma}. To be more precise, the current classical Ising annealers are implemented on several CPUs or GPUs, and the optimal result is selected from the independent and parallelized optimization processes \cite{Kowalsky2022xorsat, Zhu2015space,Aramon2019fujitsu}.

For further speedup, simulated bifurcation algorithm \cite{Goto2019sb, Tatsumura2021sb, Goto2021dsbm} and graph decoupling algorithm \cite{Aadit2022massive} have been introduced for the simultaneous updating of the spins, which are also adapted to generic spin connections and non-binary coupling coefficients. However, the simulated bifurcation algorithm focuses on the Ising problems without local fields, while the graph decoupling algorithm introduces extra heuristic procedures for isolating the spins.

In this paper, we propose an Ising annealer implemented with a parallel algorithm, which is compatible with general types of spin connections and fixed-point coupling coefficients. This annealer is named as Parallel Hamiltonian Monte Carlo Ising Annealer, or PHIA. This annealer samples feasible configurations with a highly parallelized Hamiltonian Monte Carlo (HMC) method, which derives the candidates by sliding the sampling points on the energy landscape guided by the Hamiltonian equations. HMC \cite{Neal2012hmc} is proposed as an alternative for Gibbs sampler on continuous variables. The Gaussian augmented HMC (GAHMC) was introduced in \cite{Pakman2013auxi} to extend the vanilla HMC sampling to discrete variables. However, the gradient descent method has not yet been applied to GAHMC, which severely limits its efficiency. Therefore, we derive an extended GAHMC method to solve the Ising annealing problems with an approximate gradient-based method based on our previous research \cite{li2023simulated}. In our PHIA, the updating of a position variable is only dependent on its corresponding momentum variable, thereby guaranteeing the parallelized updating of all dimensions. The approximate gradient-based approach is then introduced into the annealing algorithm after the fusion of discrete Ising problem and the continuous HMC sampler for a further speedup. Finally, the max degree of parallelism is achieved by implementing the gradient-based HMC sampling procedure on a low-cost commercial FPGA board which can simulate up to hundreds of spins. We compare the PHIA with the state-of-the-art algorithms and hardwares, including the simulated CIM \cite{Tiunov2019simcim}, D-Wave 2000Q (DW2Q) and CIM \cite{Hamerly2019benchmark,Goto2021dsbm} on benchmark problems. The experiment results show that PHIA can achieve superb performance regardless of the problem type.

\section{Results}\label{sec2}
\subsection{Parallel anealer based on HMC}\label{subsec2_1}
\begin{figure*}[!htp]
	\centering
	\includegraphics[width=1\linewidth, trim=0 110 25 100,clip]{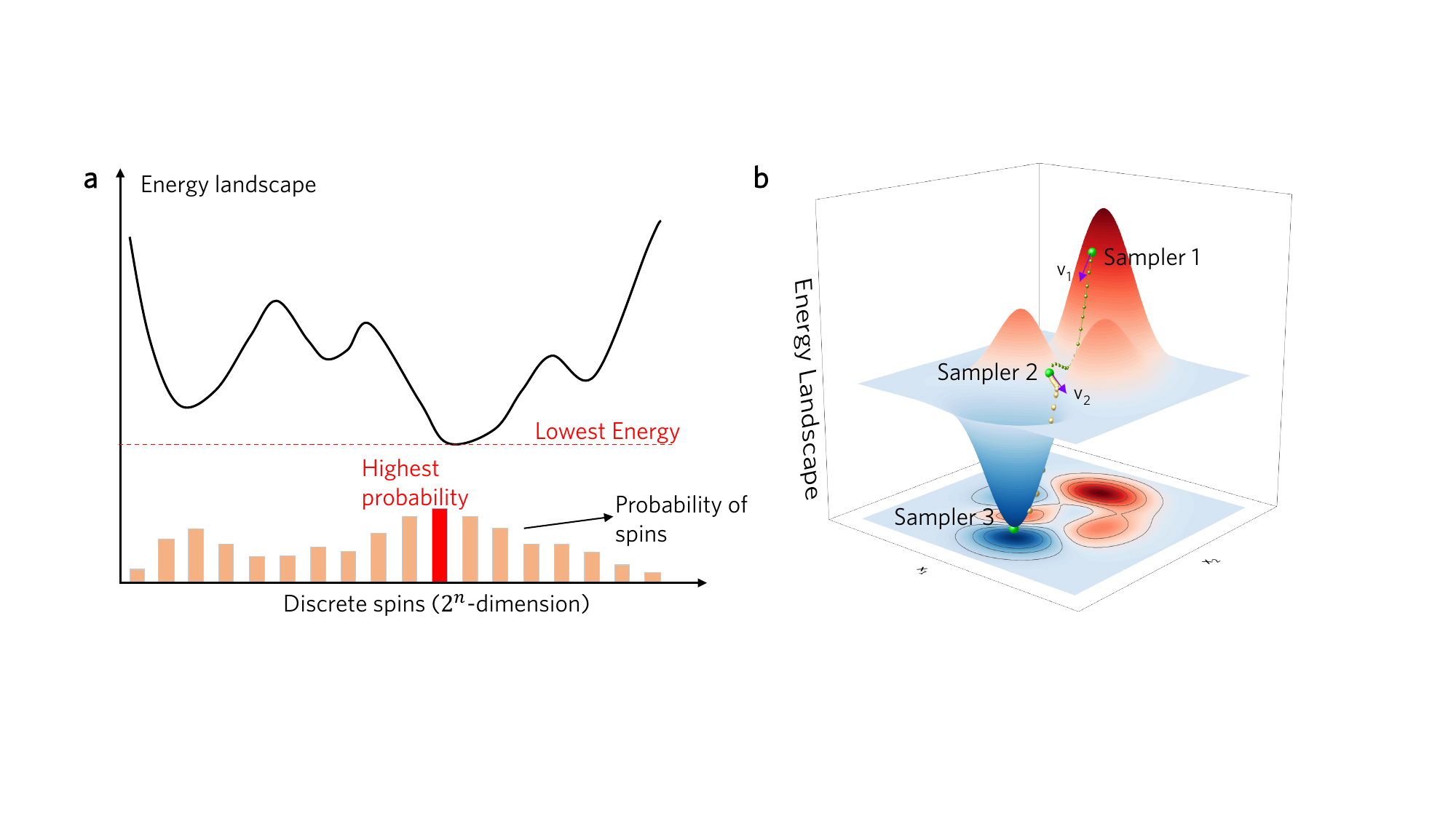}%
	\caption{
        \label{fig:theory}
        (\textbf{a}) The relation between the energy landscape of the Ising problem and the probability of spin configurations, where a low energy signifies a high sampling probability according to the Boltzmann distribution. 
        (\textbf{b}) HMC method allows a particle to slide on the energy landscape. The final location of the particle is the candidate sampling point and the initial position of the next round of sampling.}
\end{figure*}

\noindent
The proposed PHIA is used to solve the combinatorial optimization problem. Like other Ising annealers, the optimization problem reduces to seeking for the ground states of an Ising Hamiltonian \cite{Lucas2014Ising}. 

The probability of obtaining a spin configuration $\bm{s} \in \left\{-1, 1\right\}^n$ of an ensemble of $n$ spins in the Ising model follows the Boltzmann distribution
\begin{equation}
    p(\bm{s}) = \frac{1}{Z} \exp\left[-\beta E(\bm{s})\right], 
\end{equation}
where $E(\bm{s})$ is the energy of the configuration with the normalization factor $Z$, and $\beta$ is the inverse temperature. The relation between the energy landscape and the probability of spin configurations is shown in \fig{fig:theory}a. The energy landscape in Ising model is defined by the following Hamiltonian
\begin{equation}
	E(\bm{s}) = -\sum_{i<j}J_{ij}s_is_j-\sum_{i=1}^{n}h_is_i, \label{eq:ising}
\end{equation}
where the spins are mutually coupled through the internal field $J_{ij}\in\mathbb{R}$ and simultaneously affected by the local field $h_i\in\mathbb{R}$. The global minimum of \eq{eq:ising} is the solution to the optimization problem.

The traditional Gibbs sampling method samples from the Ising Hamiltonian with an acceptance probability and updates only one spin in a single run, which severely limits its scalability in high-dimension problems. In contrast, as shown in \fig{fig:theory}b, HMC is a fundamentally different approach which proposes the candidate sampling points by evolving the Hamiltonian equations on the energy landscape and updating the variables of all dimensions in parallel. More specifically, the sampling point slides on the energy landscape with randomly initialized momentum $\bm{v}$ at a certain position $\bm{x}$, governed by the Hamiltonian equations
\begin{equation}
\label{Hamiltonian equation-gen}
	\left\{
		\begin{aligned}
			\dot{x}_i&=\frac{\partial H}{\partial v_i}, \\
			\dot{v}_i&=-\frac{\partial H}{\partial x_i},
		\end{aligned}
	\right.
\end{equation}
with $H(\bm{x}, \bm{v}) = U(\bm{x}) + K(\bm{v})$. Here $U(\bm{x})$ denotes the potential energy which corresponds to $E(\bm{s})$ in the Ising model, and $K(\bm{v})=\frac{1}{2}\left\Vert \bm{v} \right\Vert^2_2$ is the kinetic energy. The Hamiltonian equations guarantee that in PHIA each pair of $x_i$ and $v_i$ can be updated in parallel.

The key result of this paper is the fusion of the gradient descent method for accelerating the HMC sampling process, based on the model proposed in \cite{Pakman2013auxi} that converts the sampling of binary variables into the sampling of continuous variables for solving the Ising optimization problems. This is done by correlating the spin variables with the continuous position variables by a joint probability distribution $p(\bm s, \bm x)$. By assuming that $U(\bm x)$, which is the continuous counterpart of $E(\bm{s})$, also obeys the Boltzmann distribution $p(\bm x)\propto \exp\left[-\beta U(\bm x)\right]$, we can derive the explicit form of $H$ as 
\begin{equation}
\label{H}
    H(\bm x, \bm v) = \beta E({\rm sgn}~{\bm x}) + \frac{1}{2}{\bm x}^T{\bm x} + \frac{1}{2}{\bm v}^T{\bm v}.
\end{equation}
More details about this derivation can be found in Appendix~\ref{ap1}. Then the Hamiltonian equations are calculated as
\begin{equation}
\label{Hamiltonian equation}
	\left\{
		\begin{aligned}
			\dot{x}_i&= v_i, \\
			\dot{v}_i&= \beta\frac{{\rm d(sgn}~x_i)}{{\rm d} x_i} I({\rm sgn}~x_i) - x_i,
		\end{aligned}
	\right.
\end{equation}
with
\begin{equation}
    \label{quasi gradient}
    I({\rm sgn}~x_i)=-\left.\frac{\partial E(\bm s)}{\partial s_i}\right\vert_{{\bm s}={\rm sgn}~\bm x}=\sum_{j=1}^{n}J_{ij}({\rm sgn}~x_j)+h_i.
\end{equation}
The exact solution to the Hamiltonian equations can be obtained by considering the zero-cross of the continuous variables \cite{Pakman2013auxi}, whereas an iterative gradient descent `leapfrog' method \cite{Neal2012hmc} will be more friendly for parallel implementation on the FPGA. However in our case, since ${\rm sgn}~x_i$ is non-differentiable, we have to invoke the following approximation
\begin{gather}
        {\rm sgn}~{x_i}\approx{\rm Tanh}(\gamma x_i), \\
		\frac{\rm d}{{\rm d}x_i}{\rm Tanh}(\gamma x_i)=\gamma [1-{\rm Tanh}^2(\gamma x_i)],	\label{eq:de}
\end{gather}
with a hyper-parameter $\gamma$ for the gradient calculation. The HMC sampling gives the current estimate of $\bm x^*$ minimizing $U(\bm x)$. Therefore, the corresponding spin configuration $\bm s^*$ calculated by ${\bm s}^* = {\rm sgn}~{\bm x}^*$ is the current estimate that minimizes $E(\bm{s})$. The ground states of the Ising Hamiltonian can be found by repeating this sampling process.

\subsection{The FPGA implementation of the parallel annealer}\label{subsec2_2}
\begin{figure*}[!htp]
	\centering
	\includegraphics[width=1\linewidth,clip]{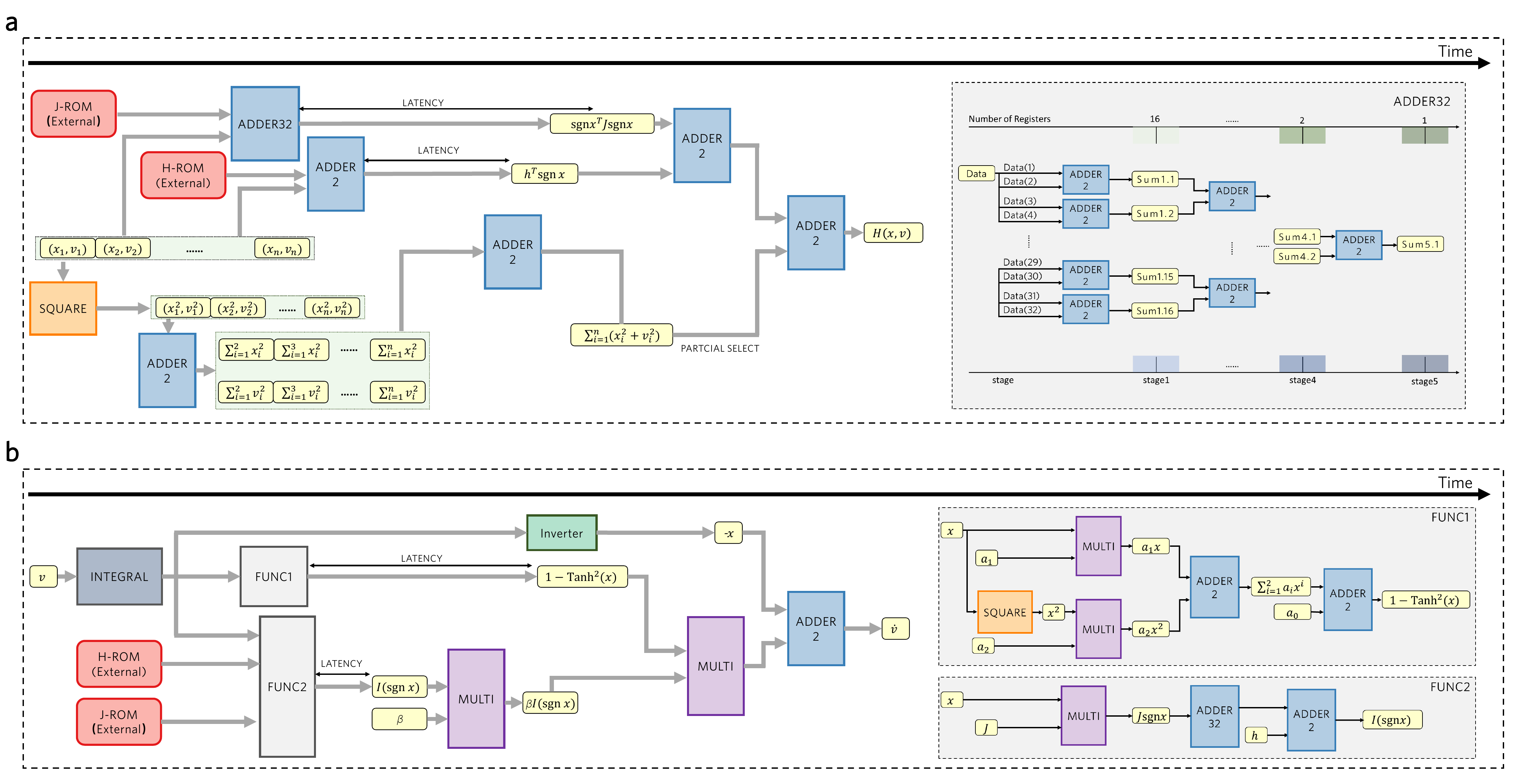}
	\caption{\label{fig:flowchart} Schematic of two main blocks in the FPGA implementation. (\textbf{a}) The block that calculates $H(\bm x, \bm v)$. ADDER32 is a 5-stage pipelined logical block composed of 31 ADDER2s. (\textbf{b}) The block that calculates $\dot{\bm v}$. The main latency originates from computing the derivative of the sign function and \eq{quasi gradient}.}
\end{figure*}

\begin{table}
    \caption{Main Logic Resources on the ALINX-AX7Z100 \label{table:resource}}
    \begin{tabular}{ccccc}
    \hline
    Device & Slice Registers&LUT&
      DSP&LUTRAM\\ 
    \hline
    ALINX-AX7Z100 &
      554,800 &
      277,400 &
      2020 &
      108,200 \\ 
    \hline
    \end{tabular}
\end{table}

\noindent
FPGA \cite{kilts2007advanced} is an integrated circuit with configurable logic blocks that admit parallel execution. In this paper, ALINX-AX7Z100 FPGA board is used for demonstration. Table \ref{table:resource} shows the characteristics of ALINX-AX7Z100. The implementation is based on the fixed-point arithmetic, which extends the solvable problems to Ising Models with non-integer $J_{ij}$ and $h_i$. 

Speed and area are the main considerations in the logic design \cite{kocc1999cryptographic}. In our implementation, increased logic resources are used for the loop unrolling in each iteration to accelerate the annealing process. In addition, at the cost of larger circuit size, pipeline is used to improve parallelism, and sharing blocks between different operations reduces the consumption of time and resources. There are two main shared blocks which calculate \eq{H} and \eq{Hamiltonian equation} as shown in \fig{fig:flowchart}a and \fig{fig:flowchart}b, respectively. The parameters for the internal field $J_{ij}$ and local field $h_i$ are stored in Read Only Memory (ROM). In \fig{fig:flowchart}b, FUNC1 and FUNC2 can work in parallel as there is no data dependency between them. With the pipeline design, the updating of $\bm{x}$ and $\bm{v}$, which is the most time-consuming part of the calculation, can be significantly shortened. Note that we adopt a quadratic polynomial approximation to \eq{eq:de} which only takes 2 MULTIs, 2 ADDER2s and 1 SQUARE. Additional control circuitry such as finite state machine (Appendix~\ref{ap2}) is required to organize the logic blocks.

\section{Numerical results}\label{sec3}
\subsection{Problem definition for benchmark experiments}\label{subsec3_1}
Next, we present a number of benchmark experiments in which different types of combinatorial optimization problems are considered. The definition of these problems is summarized in Table~\ref{tab:data}. The maxcut problems, including maxcut\_d3 and maxcut\_dense, are aimed at finding the optimal partition of a given unweighted graph \cite{Hamerly2019benchmark}. The sk problems, including sk\_bool, sk\_ising, and sk\_uniform, are aimed at computing the ground states of the Sherrington-Kirkpartrick spin glass model with different types of coupling coefficients \cite{SK1975spin, Hamerly2019benchmark}. The nae\_3\_sat problem is a variant of the satisfiability problem \cite{oshi2022qubo}. The spin\_model is an Ising problem whose coupling coefficients are randomly chosen from the uniform distribution. In particular, coupling coefficients in both sk\_uniform and spin\_model are float-point numbers, while coefficients in the other problems are not. 

\begin{table}[!htp]
  \caption{Datasets and Parameters for Benchmarking \label{tab:data}}
  \begin{tabular}{@{}ccc@{}}
  \toprule 
  problem & internal $J_{ij}$ & external $h_i$ \\
  \midrule
  maxcut\_dense & $\{0, 1\}$ & $0$ \\
  maxcut\_d3 & $\{0, 1\}$ & $0$ \\
  sk\_bool & $\{0, 1\}$ & $0$ \\
  sk\_ising & $\{-1, +1\}$ & $0$ \\
  sk\_uniform & $0$ or $U(0, 1)$ & $0$ \\
  nae\_3\_sat & $\{0, \pm 1, \pm 2, \cdots , \}$ & $0$ \\
  spin\_model & $U(-1, 1)$ & $U(-1, 1)$ \\
  \botrule
  \end{tabular}
\end{table}

\begin{figure*}[!htp]
	\centering
	\includegraphics[width=0.9\linewidth, trim=0 0 40 0, clip]{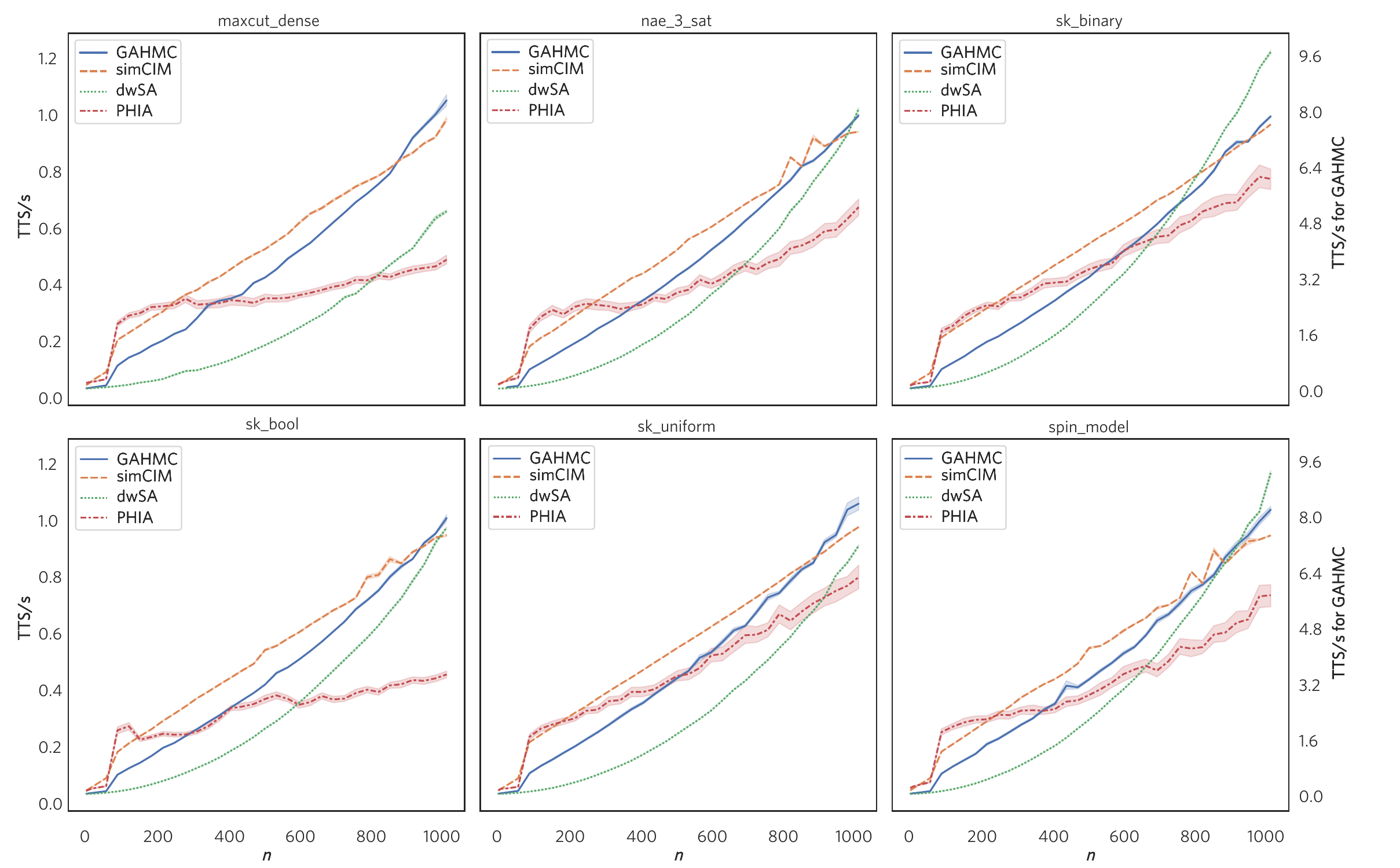}%
	\caption{\label{fig:sim}The simulation results on different Ising annealers. With the dimension of the problem grows, the PHIA outperforms the other Ising annealers on all types of problems. The TTSs of the GAHMC are much larger than the other annealers, and thus are indicated with a different axis on the right with a different scale.}
\end{figure*}

\subsection{The simulation of the parallel annealer}\label{subsec3_2}
Ising annealers in general do not guarantee that the global optimum can be obtained. Hence, we employ the Time-To-Solution (TTS) as the performance index to quantify the expected time to attain the best solution with a ninety nine percent confidence \cite{king2015benchmarking}. The optimal solution is estimated using the third-party Python library `dwave-neal' (dwSA) \cite{dwaveneal}. TTS is defined by
\begin{equation}
{\rm TTS}=T_1\left[\frac{\lg(1-0.99)}{\lg(1-P)}\right],
\end{equation}
where $T_1$ is the time of a single run, and $P$ is the probability of finding the ground state within a single annealing. \fig{fig:sim} depicts the average time to approach the optimal solutions for various combinatorial optimization problems as the dimension of problems grows.

\begin{figure*}[t]
	\centering
	\includegraphics[width=0.9\linewidth,clip]{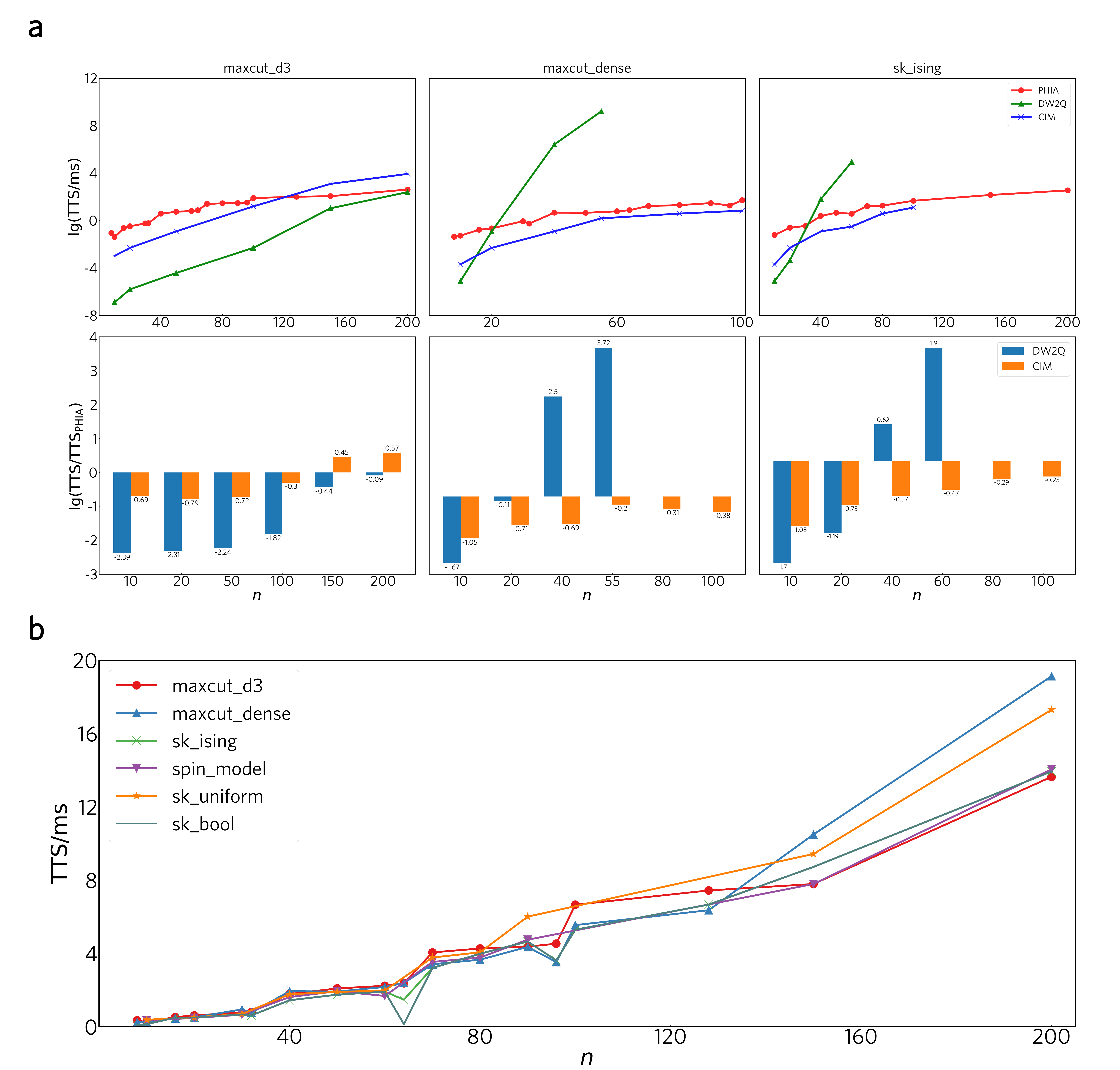}
	\caption{\label{fig:hardware_experiment}Data of CIM and DW2Q are borrowed from \cite{Hamerly2019benchmark}. (\textbf{a}) The comparison of TTS on 3 types of problems. $\rm TTS_{PHIA}$ is the TTS of PHIA. (\textbf{b}) Performance of PHIA on various combinatorial optimization problems.}
\end{figure*}

In our experiments, all types of optimization problems are firstly converted into Ising annealing problems. In each type of problem, 100 instances with dimension that scales from 8 to 1024 are tested to evaluate the scalability of four annealers on the same CPU (AMD Ryzen 7 5800H at 3.20 GHz with 16-GB memory) in Python. The simulated annealer `dwSA' is implemented by directly calling the API provided by `dwave-neal', and the simulations of coherent Ising machine `simCIM' and `GAHMC' are implemented using Python according to the algorithm designs in \cite{Pakman2013auxi}. As shown in \fig{fig:sim}, the PHIA outperforms the other three annealers on all six types of problems when the problem size increases beyond 800. In particular, it can be observed that the TTS of GAHMC is at least 8 times larger than the other annealers, which demonstrates the huge advantage of our proposed PHIA over the vanilla GAHMC without using the gradient descent. Moreover, dwSA surges exponentially and the simCIM rises linearly as the dimension increases. In contrast, the PHIA obeys a sub-linear growth law, which demonstrates its superb scalability in high dimensions.

\subsection{Demonstration of the parallel annealer on FPGA}\label{subsec3_3}
There exists a significant difference between fraction and integer calculation in the logic circuit design, where integer calculation often consumes more resources and time. The benchmark problems listed in Table~\ref{tab:data} can be divided into integer ones and fractional ones depending on whether the internal $J_{ij}$ and external $h_i$ are integers. PHIA can deal with both problems.

We compare the FPGA-based PHIA against DW2Q and CIM in \fig{fig:hardware_experiment}a. It should be noted that DW2Q and CIM are specifically designed for optimization problems with integer coefficients, and thus we are constraining the capability of our device in order to make a comparison on the small subset of problems that can be solved using the FPGA-based PHIA. CIM and DW2Q have shown that they can be faster than the classical counterparts, depending on the type of the problem. Due to the hardware constraints, they can also be slower. For example, the performance of DW2Q is not satisfying on maxcut problem with dense connections, since the hardware does not directly support the implementation of dense connections. In contrast, the PHIA maintains a stable sub-linear growth in running time for different problem types. In Appendix~\ref{ap2}, we have given an estimate of the time for a single run on the FPGA, which clearly shows that the running time of PHIA will not grow exponentially with the problem size. Note that CIM has achieved similar performance, at the expense of complex optical equipment.

In addition to the integer problems, the PHIA achieves stable and satisfying performance on the fractional problems, which is demonstrated in \fig{fig:hardware_experiment}b.

\section{Discussion and conclusion}\label{sec4}
We have proposed a scalable heuristic annealing algorithm based on HMC sampling for solving the Ising problems. The gradient-based approach has been fused into the previous plan which introduces significant acceleration and facilitates parallel processing. Due to the simultaneous updating of each dimension, we have demonstrated a significant increase in scalability and built a prototype annealer that solves Ising problems with up to 200 spins on a single FPGA.

The proposed annealer outperforms its counterparts in various benchmark problems as the problem size increases. When implemented with a single FPGA, the annealer also achieves a satisfying performance when compared with the state-of-the-art hardware CIM and DW2Q on optimization problems with integer coefficients. In addition, the proposed hardware is adapted to both integer and fraction problems. Therefore, this annealer could serve as the benchmark to measure the performance of quantum hardware. 

Due to the sub-linear growth of the execution time of the PHIA, it is promising to consider multi-FPGA systems for developing an ultrafast solver that is compatible with Ising problems of thousands of spins, which corresponds to large-scale combinatorial optimization problems that is still challenging in classical computing.

\begin{appendices}
\setcounter{equation}{0}

\renewcommand\theequation{A\arabic{equation}} 

\section{Implementation details of the Ising annealer via the discrete HMC sampler\label{ap1}}
In HMC, the physically inspired Hamiltonian is defined using the continuous variables $\bm x$ and $\bm v$. HMC guarantees the conservation of total energy and an acceptance probability of 1 for the sampling points \cite{Neal2012hmc}, while the probability of rejection in the traditional Monte Carlo sampling is nonzero.

However, the energy function in \eq{eq:ising} is defined over binary variables which are incompatible with the continuous HMC. Here we have used the approach proposed in \cite{Pakman2013auxi} to convert the sampling of binary spins into the sampling of continuous distributions, which is detailed as follows.

Assume the joint probability distribution of the continuous variables $\bm x$ and the discrete variable $\bm s$ is written as 
\begin{equation}
    p(\bm s, \bm x) = p(\bm s) p(\bm x \mid \bm s).
\end{equation}
The conditional probability is set as \cite{Pakman2013auxi}
\begin{equation}
	p(\bm{x}\mid\bm{s})=g(\bm{x})\delta({\rm sgn}~\bm{x}-\bm{s}),\label{eq:s_condi}
\end{equation}
with $g(\bm{x})$ being a continuous function that can be arbitrarily chosen. It can be proved that $g(\bm{x})$ is a probability distribution. Therefore, a trivial choice for $g(\bm{x})$ is
\begin{equation}
    g(\bm x)={(2/\pi)^{n/2}}\exp\left(-\frac{1}{2}{\bm x}^T{\bm x}\right).
\end{equation}
In this case, the marginal probability distribution $p(\bm x)$ can be derived as
\begin{equation}
		p(\bm{x})=p({\bm s}={\rm sgn}~\bm{x})g(\bm{x}).\label{eq:s_px}
\end{equation}
According to the Boltzmann distribution $p(\bm x)\propto \exp\left[-\beta U(\bm x)\right]$, the potential energy is given by
\begin{equation}
	\begin{aligned}
		\beta U(\bm{x}) &= -\ln \left[p(\bm{x})\right] \\
		&= -\ln \left[p(\bm s={\rm sgn}~\bm{x})\right] -\ln \left[g(\bm{x})\right] \\ 
        &= \beta E({\rm sgn}~{\bm x}) +\frac{1}{2}{\bm x}^T{\bm x} + C(\beta), \label{eq:s_ux}
	\end{aligned}
\end{equation}
with $C(\beta)$ being a constant. Since a constant factor does not affect the value of $\bm{x}$ that minimizes $U(\bm x)$, the Hamiltonian $H(\bm x, \bm v)$ can be formulated as 
\begin{equation}
    H(\bm x, \bm v) = \beta E({\rm sgn}~{\bm x}) + \frac{1}{2}{\bm x}^T{\bm x} + \frac{1}{2}{\bm v}^T{\bm v}.
\end{equation}
Then the sampling on $U(\bm x)$ can be conducted by simulating the Hamiltonian equations \eq{Hamiltonian equation}, using the Leapfrog numerical integration method as outlined in \cite{Neal2012hmc}. In particular, $\bm{x}$ and $\bm{v}$ are updated by using an approach similar to the expectation-maximization (EM) algorithm
\begin{equation}
\label{equation:EM}
	\left\{
		\begin{array}{lcl}
			\bm{x}^{(i+1)} & = & \bm{x}^{(i)} + \epsilon \bm{v}^{(i)}, \\
			\bm{v}^{(i+1)} & = & \bm{v}^{(i)} - \epsilon \nabla H(\bm{x}^{(i+1)}), \\
		\end{array}
	\right.
\end{equation}
where $\epsilon$ controls the step size of each update, and $\nabla H(\bm x)$ denotes the gradient of $H$ in the $\bm x$ direction
\begin{equation}
    \nabla H(\bm x) = \frac{\partial H}{\partial \bm x}.
\end{equation}

\section{Estimation of the time for a single run on the FPGA}
\label{ap2}

\noindent
The global clock frequency of the FPGA board is 100~MHz. The annealing process is planted into the FPGA as a state machine of 5 states. The states are listed in Table~\ref{tab:states}.
\begin{table}
  \caption{The behaviors of states.}\label{tab:states}
  \begin{tabular}{@{}cl@{}}
  \toprule
  States & Behavior \\
  \midrule
  1 & Initialization\\
  2 & Updating $v$ in the first iteration\\
  3 & Executing the rest iterations of the EM algorithm\\
  4 & Calculating the acceptance rate and adjusting\\& the temperature of the annealing algorithm \\
  5 & Determining whether the PHIA has obtained the \\& optimal solution and saving the best result achieved \\& so far\\
  \botrule
  \end{tabular}
\end{table}

One clock cycle ($\rm Clk$) serves as the unit to measure the time taken by each state
\begin{equation}  
{\rm Clk}=\frac{1}{100~{\rm MHz}}=10~{\rm ns}.      
\end{equation}
The time cost by the State 1 is approximately
\begin{equation}
    T_{s1}=(\frac{n}{2}+5)~{\rm Clk}.
\end{equation}
The time cost by the State 2 is about
\begin{equation}
    T_{s2}=(\frac{n}{2}+10+(n+2)\lfloor\frac{n}{32}\rfloor)~{\rm Clk}
\end{equation}
The time for a single run is mainly consumed in repeating the gradient update procedures for ${L}$ times in State 3, which is denoted by $T_{s3}$. $T_{\rm{d}}$ is the latency caused by the Block shown in \fig{fig:flowchart}b, which constitutes the major part of $T_{s3}$. $T_{\rm{d}}$ is calculated by
\begin{equation}
    T_{\rm{d}}=(n+12+(n+2)\lfloor\frac{n}{32}\rfloor)~{\rm Clk},
\end{equation}
where $\lfloor \cdot \rfloor$ denotes the Greatest Integer Function. Then $T_{s3}$ is given by
\begin{equation}
    \begin{aligned}
    T_{s3}&=LT_{\rm{d}}+L(13+n)\\
    &=L(2n+25+(n+2)\lfloor\frac{n}{32}\rfloor)~{\rm Clk}.   
    \end{aligned}
\end{equation}
The time for a single run is estimated as 
\begin{equation}
    \begin{aligned}
     {T_{\rm est}}&=T_{s1}+T_{s2}+T_{s3}\\
    &=((2L+1)n+15+25L+  \\&(L+1)((n+2)\lfloor\frac{n}{32}\rfloor))~{\rm Clk},
      \label{eq:tts_single_est}
    \end{aligned}
\end{equation}
which clearly indicates that the running time will not grow exponentially with the problem size $n$.

\end{appendices}


\section*{Declarations}

\subsection*{Funding}

\noindent This work was supported by the National Key R\&D Program of China (Nos. 2022YFB3304700 and 2022YFB3103100) and the National Natural Science Foundation of China (Nos. 62173296 and 62273154).

\hfill

\subsection*{Availability of data and materials}

\noindent The data and materials that support the findings of this study are available from the corresponding author upon reasonable request.

\hfill

\bibliography{sn-bibliography}

\end{document}